\documentclass[a4paper,aps,nofootinbib]{revtex4}                  
\usepackage{graphicx}
\usepackage{amsmath} 
\usepackage{amssymb}                                                                   
                                                                  
\def\beq{\begin{equation}}                                        
\def\eeq{\end{equation}}                                          
                                                                  
\def\beqn{\begin{eqnarray}}                                       
\def\eeqn{\end{eqnarray}}                                         
                                                                  
\begin{document}                                                  
                                                                  
\title{Kalb-Ramond field and Dirac-K\"ahler equation}

\author{V. A. Pletyukhov}                                         
\affiliation{Brest State University, Brest, Belarus}              
\author{V. I. Strazhev}                                           
\affiliation{Belarusian State University, Minsk, Belarus}         

\begin{abstract}
Matrix and tensor formulations of the relativistic wave equation
providing description both an electromagnetic field (photon) and a
massless Kalb-Ramond field with the zero helicity (notoph) are
given. It is shown that this equation is a particular case of the
Dirac-K\"ahler system.
\end{abstract}

\maketitle

%%%%%%%%%%%%%%%%%%%%%%%%%%%%%%%%%%%%%%%%%%%%
%% MAINMATTER
%%%%%%%%%%%%%%%%%%%%%%%%%%%%%%%%%%%%%%%%%%%%

\section{Introduction}

It was demonstrated by Ogievetski and Polubarinov [1] in 1966 that if we use
the tensor-potential $\Phi_{\mu \nu}~(\Phi _{\mu \nu}=-\Phi_{ \nu\mu} $) to
describe massless field then corresponding massless particle has a single
polarization state associated with zero helicity. This particle was called
by the authors [1] as "notoph".

When in the theory of the photon the potential and the field intensity are
represented by the four-vector and the second-rank antisymmetric  tensor,
respectively, whereas in the theory of the notoph
the opposite statement is true, namely the
antisymmetric tensor corresponds to the potential,
the four-vector corresponds to the
field intensity. The last statement needs some explanation. In fact, the
notoph field intensity is given by a completely antisymmetric third-rank
tensor, however, its components are like the components of the axial four-vector
under the Lorentz group transformations.

In the authors'[1] opinion, the notoph is not a scalar particle with zero mass
and helicity. Really, if the notoph is converted to a pair, the total angular
momentum of the pair equals unity. The authors [1] came to the conclusion
also that the notoph could not be associated with well-known interactions.

Kalb and Ramond [2] practically rediscovered in 1974 the notoph on the basis
of another approach. The definition``Kalb-Ramond field'' is now became
established. To describe interactions of a charged particle represented in
4-dimensional space-time as a closed string, they suggested using  a
tensor-potential $\Phi_{\mu \nu}$. Unlike Ogievetski and Polubarinov, these
authors have considered a massless particle with one degree of freedom
described by the tensor-potential $\Phi_{\mu \nu}$ as a spin-0 massless
particle (massless scalar meson). The theory of Kalb -- Ramond field
(notoph) is discussed in different aspects in many publications (see for
example [3-6] and references cited there). In the present paper we propose
a new approach to this subject based on the theory
of relativistic wave equations (RWE) [7-9].

\section{Photon and notoph}

It is well known that the photon possessing two polarization states
(helicity $\pm $ 1) is described by the vector-potential $A_\mu $ obeying
the second-order equation
\beq
\label{eq1}
\Box A_\mu -\partial _\mu \partial _v A_v=- j_\mu ,
\eeq
where $j_\mu$   is electric current.

Equation (\ref{eq1}) is invariant with respect to the gauge transformations
\beq
A_\mu \to A_\mu  + \partial _\mu \Lambda \left( x \right).
\label{eq2}
\eeq
The field intensity, invariant with respect to transformations (\ref{eq2}),
corresponds to the second-rank antisymmetric tensor
\begin{equation}
F_{\mu  v}= \partial _\mu A_v  - \partial _v A_\mu.
\label{eq3}
\end{equation}
A massless vector field can be also described by the tensor-potential
$\Phi_{\mu  v}  \left( {\Phi _{\mu v} = - \Phi _{v \mu}}
\right)$ that satisfies equation~(1)
\begin{equation}
\label{eq4}
\Box\Phi _{\mu  v} - \partial _\mu \partial _\lambda \Phi _{\lambda
v} + \partial _v \partial _\lambda \Phi _{\lambda\mu }  =
 - j_{\mu v},
\end{equation}
where $j_{\mu v} $ is the tensor current $\left( {j_{\mu v} =-
j_{v\mu },~\partial _\mu j_{\mu v}=0} \right).$
Equation (\ref{eq4}) is invariant with respect to the gauge transformations
\begin{equation}
\label{eq5}
\Phi _{\mu v} \to \Phi _{\mu v}  + \partial _\mu \Lambda _v
\left( x \right) - \partial _v \Lambda _\mu \left( x \right).
\end{equation}
One can choose a gauge under which an interaction-free equation corresponding
to equation (\ref{eq4}) takes the following form
\beqn
\label{eq6}
\Box\Phi _{\mu v}=0,
\\
\label{eq7}
\partial _\mu \Phi _{\mu v}=0.
\eeqn
Equations (\ref{eq6}), (\ref{eq7}) still remain invariant with respect to the gauge
transformations of (\ref{eq5}) on the condition
of the following restriction of the
gauge functions
\beq
\label{eq8}
\Box\Lambda _\mu - \partial _\mu \partial _v \Lambda _v =0.
\end{equation}
The gauge transformations (\ref{eq5}), (\ref{eq8})
and additional requirements (\ref{eq7}) lead to
one independent component of the tensor-potential $\Phi _{\mu v} $. This
result is obtained in the quantum case as well. Note, however, requirements
(\ref{eq7}) are imposed in the quantum case on physical state vectors. So, a
massless particle described by equations (\ref{eq6}) and (\ref{eq7})
has  one polarization
state associated with zero helicity.

The  third-rank antisymmetric tensor $F_{\mu v\lambda}  $
connected with the tensor-potential $\Phi
_{\mu v} $ by the relation
\begin{equation}
\label{eq9}
F_{\mu v\alpha}  =\partial _{\mu}  \Phi _{v,\alpha}
+ \partial _{\alpha}  \Phi _{\mu v}+\partial _{v} \Phi_{\alpha \mu},
\end{equation}
is considered as a Kalb-Ramond field intensity.

The description of the interaction arising in the case of open strings (the
string ends considered as point charged particles of opposite sign)
involves, together with the potential $\Phi _{\mu v} $, an ordinary
4-dimensional vector-potential. In other words, to describe the interaction
of open strings in 4- dimensional space-time, one needs two fields: an
ordinary electromagnetic field (having two polarization states with the helicity
$\pm $ 1) and the Kalb-Ramond field (with the helicity 0).

Now let us compare the first-order tensor formulation of the Kalb-Ramond
field (notoph) to the formulation of the electromagnetic field
(photon). Maxwell's equations may be written in the following form
\beq
\begin{split}
\partial _{v} F_{\mu v} = j_{\mu},
   \\
- \partial _{\mu}  A_{v}  + \partial _{v} A_{\mu}+F_{\mu v} = 0.
\end{split}
\label{eq10}
\eeq
The equations of the Kalb-Ramond field are of the form [3]
\beq
\begin{split}
\partial _{\mu}  F_{\mu v\alpha}  = j_{v\alpha},
 \\
 {\partial _{\mu}  \Phi _{v\alpha}   + \partial _{\alpha}
\Phi _{\mu  v}  + \partial _{v} \Phi _{\alpha  \mu}     +
F_{\mu v\alpha}  = 0}.
\end{split}
\label{eq11}
\eeq
System (\ref{eq10}) is invariant with respect to gauge transformations (2) and
system (\ref{eq11}) is invariant with respect
to gauge transformations (\ref{eq5}).
Here $F_{\mu v} $ and $F_{\mu v\alpha}  $ are field intensities; $A_{\mu
} $ and $\Phi _{\mu v} $ are potentials of the considered field systems. In
the case of open strings the source $j_{\mu v} $ describes the current
created by the string ``body'', $j_{\mu}  $ is the current caused by the
string ends. They are connected by the following formula [2]:
\begin{equation}
\label{eq12}
j_{\alpha}  = \partial _{\mu}  j_{\mu \alpha}.
\end{equation}

\section{ Relativistic wave equations and massless vector field}

An analysis of systems (\ref{eq10}) and (\ref{eq11}) may be achieved by writing them in the
matrix form of the first-order relativistic wave equation (RWE). In the
interaction-free case we have
\begin{equation}
\label{eq13}
\left( {\gamma _\mu \partial _\mu + \gamma _0}
\right)\psi  = 0.
\end{equation}

Under the Lorentz group  a wave function $\psi $
transforms as some set of irreducible representations of this group
forming what is known as linking scheme. The matrix $\gamma_0 $ may be
proportional to the unity matrix
\begin{equation}
\label{eq14}
\gamma _0 =\kappa I,
\end{equation}
where $\kappa $ is parameter with a dimension of mass $\left( {\hbar  =
c = 1} \right)$. In this case equation (13) describes a massive
particle. Should the matrix $\gamma _{0} $ be a singular (projective) one,
null matrix including, equation (13) gives description for a massless
particle (see [9]). The matrix $\gamma _{4} $ plays  the main role in
determining the spin structure of the field described by equation (13). In
the Gel'fand- Yaglom basis [8] it has the form of a direct sum of so called
spin blocks
\beq
\gamma _4 =  \oplus \sum\limits_s {C^s\times I_{2s +1} }.
\label{eq15}
\eeq

The spin block $C^{s}$ in (15) is constructed from the elements $C_{\tau
 {\tau} '}^{s} $, where $\tau \sim \left( {l_{1} ,l_{2}}
\right)$ and ${\tau} '  \sim \left( {l_{1} ^{\prime} ,l_{2}
^{\prime}}  \right)$ are linking irreducible Lorentz group representations,
i.e. such representations for which $l_{1} ^{\prime}  = l_{1}
\pm {\frac{{1}}{{2}}},~l_{2} ^{\prime}  = l_{2}
\pm {\frac{{1}}{{2}}}$. In such a case, the block $C^{s}$ includes the
elements $C_{\tau   {\tau} '}^{s}  $ corresponding only to the linking
representations $\tau $ and ${\tau} '$ with following restrictions:
\begin{equation}
\label{eq16}
\left| {l_1 - l_2 } \right|\le S \le \left| {l_1+l_2 }
\right|,\quad \left| {l_1 ^\prime  - l_2
^\prime } \right|\le S \le \left| {l_1 ^\prime  +l_2
^\prime } \right|.
\end{equation}
The representations satisfying the requirements of (\ref{eq16}) are regarded as
forming the spin block $C^{s}$. In the massive case, one can state the
particle has the spin $S$ when the matrix $\gamma _{4} $ (15) involves the
spin block $C^{s}$ with nonzero eigenvalues. It is clear that for the
description of the spin $S$ the linking scheme must have, at least, two
irreducible linking representations forming the block $C^{s}$. On going to
the massless case $\left( {\kappa I \to \gamma _{0} }
\right.$, where ${\left| {\gamma _{0}}  \right|} =\left. {0}
\right)$, there may be a possibility of the disappearance of the some spin
values or of spin projections existing for the massive field. However, the
appearance of new spin values, not found for a massive analog, is excluded.

Returning back to tensor systems (\ref{eq10}) and (\ref{eq11}), we can write them, with no
regard to the sources, in the form (13), where $\gamma _{\mu}  $ and
$   \gamma _{0} $ are dimensional matrices $10\times 10$ ($\gamma
_{0} $ is singular), $\psi $ is 10-component wave function.
In case (\ref{eq10}) we have
$\psi =\left(
{\begin{array}{l}
 A_\mu \\
 F_{\mu v} \\
 \end{array}} \right)$  and $\psi = \left(
{\begin{array}{l}
 F_{\mu v\alpha } \\
 \Phi _{\mu v} \\
 \end{array}} \right)$ in case (\ref{eq11}).
 The linking schemes for the Lorentz
group irreducible representations associated with systems
(\ref{eq10}) and (\ref{eq11}) are of the form
\begin{equation}
\label{eq17}
\left( {0,1} \right) - \left(
{\frac{1}{2},\frac{1}{2}} \right) -\left( {1,0}
\right),
\end{equation}
\begin{equation}
\label{eq18}
\left( {0,1} \right) -\left(
{\frac{1}{2},\frac{1}{2}} \right)^\prime  -\left(
{1,0} \right),
\end{equation}
where $\left( {\frac{1}{2},\frac{1}{2}} \right)$ is  four-vector
representation, $\left[ {\left( {0,1} \right) \oplus \left(
{1,0} \right)} \right]$ is second-rank antisymmetric tensor representation,
$\left(
{\frac{1}{2},\frac{1}{2}} \right)^\prime $ is representation of the
tensor $F_{\mu v\alpha } $ identical to  pseudo-vector representation,
sign ``--`` denotes linking. Let us denote the representations entering
into (\ref{eq17}): $\left( {\frac{1}{2},\frac{1}{2}} \right)\sim
1,\left( {0,1} \right)\sim 2,\left({1,0} \right)\sim 3$,
and into (\ref{eq18}): $\left(
{\frac{1}{2},\frac{1}{2}} \right)^\prime \sim 1,\left(
{0,1} \right)\sim 2,\left( {1,0} \right)\sim 3$.
Imposing constraints on the elements $C_{\tau {\tau }'}^s $,
following from the relativistic invariance requirement for RWE (\ref{eq13})
and from
the possibility of its derivation from the corresponding Lagrangian [8], we get
the following form of the matrix $\gamma _4 $ in the Gel'fand-Yaglom basis:
\begin{equation}
\label{eq19}
\gamma _4 =\left( {{\begin{array}{*{20}c}
 {C^0} \hfill & \hfill \\
 \hfill & {C^1\times I_3 } \hfill \\
\end{array} }} \right),
~
C^0=0,
~
C^1=\frac{1}{\sqrt 2 }\left( {{\begin{array}{*{20}c}
 0 \hfill & 1 \hfill & 1 \hfill \\
 1 \hfill & 0 \hfill & 0 \hfill \\
 1 \hfill & 0 \hfill & 0 \hfill \\
\end{array} }} \right).
\end{equation}

For the Kalb-Ramond field we have
\begin{equation}
\label{eq20}
\gamma _4 =\left( {{\begin{array}{*{20}r}
 {C^0} \hfill & \hfill \\
 \hfill & {C^1\times  I_3 } \hfill \\
\end{array} }} \right),
~
C^0 =0,
~
C^1 =\frac{1}{\sqrt 2  }\left( {{\begin{array}{*{20}c}
 {~~0} \hfill & {1} \hfill & { - 1} \hfill \\
 {~~1} \hfill & {0} \hfill & {~~0} \hfill \\
 { - 1} \hfill & {0} \hfill & {~~0} \hfill \\
\end{array} }} \right).
\end{equation}
The matrix $\gamma _0 $ in these cases is, respectively, of the forms
\beqn
\label{eq19_1}
\gamma _0=\left(
{{\begin{array}{*{20}c}
 {0_4 } \hfill & \hfill \\
 \hfill & {I_6 } \hfill \\
\end{array} }} \right)
\eeqn
and
\begin{equation}
\label{eq20_1}
\gamma _0=\left( {{\begin{array}{*{20}c}
 {I_4 } \hfill & \hfill \\
 \hfill & {O_6 } \hfill \\
\end{array} }} \right).
\end{equation}

Comparison between the matrices $\gamma _{4} $ (\ref{eq19}) and $\gamma _{4}
$ (\ref{eq20}) demonstrates that they have the same spin structure
both the Kalb-Ramond field (notoph) and the
electromagnetic field (photon). Differences in
the matrix structure of $\gamma _{0} $ for these cases lead to the
differences of possible physical states (helicity values) for the fields
under consideration.
The matrices (\ref{eq19}), (\ref{eq20}) and (\ref{eq19_1}), (\ref{eq20_1})
could not be simultaneously
transformed to each other by any unitary transformation. It means, as
evident, the physical nonequivalence of electromagnetic and Kalb-Ramond fields.

It is important to note, there exist
possibility for their simultaneous description. In [4] this problem has been
considered on the basis of the modified Bargmann-Wigner formalism for the
symmetric second-rank spinor. We propose another method of describing the
Maxwell - Kalb-Ramond field.

Let us consider the following linking scheme
\beq
\unitlength= 0.20mm
\begin{picture}(0,+50)(0,0)
\special{em:linewidth 0.4pt} \linethickness{0.4pt}
%\put(-60,-2){$(0,1) $}
\put(-70,0){$(0,1) $}
\put(+80,0){$(1,0) $.}
\put(-2,+40){$({1\over 2},{1\over 2} ) $}
\put(-2,-40){$({1 \over 2},{1\over 2} )'$}

\put(-50,+15){\line(2,+1){45}}
\put(-50,-10){\line(2,-1){45}}
\put(+90,+15){\line(-2,+1){45}}
\put(+90,-10){\line(-2,-1){45}}

\end{picture}
\label{eq21}
\vspace{1cm}
\eeq
According to the Gel'fand-Yaglom method [8] we form zero-mass RWE, where
$\psi $ is the 14-component wave function $\psi =\left(
{A_{\mu}  ,F_{\mu v\alpha}  ,\Phi _{\mu v}}  \right)$. The
matrices $\gamma _{4} $ and $\gamma _{0} $ are of the form
\begin{equation}
\label{eq22}
\gamma _4 =\left( {{\begin{array}{*{20}c}
 {C^0} \hfill & \hfill \\
 \hfill & {C^1\times I_3 } \hfill \\
\end{array} }} \right),~C^0 =\left( {O_2 }
\right),~C^1 =\frac{1}{\sqrt 2 }\left(
{{\begin{array}{*{20}c}
 0 \hfill & {~~0} \hfill & 1 \hfill & {~~1} \hfill \\
 0 \hfill & {~~0} \hfill & 1 \hfill & { - 1} \hfill \\
 1 \hfill & {~~1} \hfill & {0} \hfill & {~~0} \hfill \\
 1 \hfill & { - 1} \hfill & {0} \hfill & {~~0} \hfill \\
\end{array} }} \right),
\end{equation}
\begin{equation}
\label{eq23}
\gamma _0 = \left( {{\begin{array}{*{20}c}
 {I_8 } \hfill & \hfill \\
 \hfill & {O_6 } \hfill \\
\end{array} }} \right)= \left( {{\begin{array}{*{20}c}
 {I_2 } \hfill & \hfill & \hfill \\
 \hfill & {I_6 } \hfill & \hfill \\
 \hfill & \hfill & {O_6 } \hfill \\
\end{array} }} \right),
\end{equation}
and we use the following numbering of the corresponding representations
$\left( {{\frac{{1}}{{2}}},{\frac{{1}}{{2}}}} \right)\sim
1, \quad \left( {{\frac{{1}}{{2}}},{\frac{{1}}{{2}}}}
\right)^{\prime} \sim 2, \quad \left( {0,1}
\right)\sim 3, \quad \left( {1,0} \right)\sim
4 $ involved in (\ref{eq21}). Structure (\ref{eq22}) of the matrix $\gamma _{4} $ and
its spin blocks $C^{0},C^{1}$ demonstrate that a massless field
described by this RWE correspond to spin value 1 in the case of particle
with a mass. The eigenvalues $\pm 1$ of the spin block $C^{1}$ are
two-fold degenerate. This degeneracy is associated with the description of a
photon and the notoph on the basis of single indecomposable (irreducible) RWE.
The matrix $\gamma _{0} $ (\ref{eq23}) ``cuts out'' a half (three of six) of the
states with spin equal 1, leaving for the photon and the notoph three degrees of
freedom in total.

The tensor form of RWE (13), (\ref{eq21})--(\ref{eq23}) is following
\begin{subequations}
\label{sys1}
\beqn
\label{sys1a}
\partial _v \Phi _{\mu v} + A_\mu  = 0,
\\
\label{sys1b}
\partial _\mu \Phi _{v\alpha }  +\partial _\alpha \Phi _{\mu v}
+ \partial _v \Phi _{\lambda \mu } + F_{\mu v\alpha
} =0,
\\
\label{sys1c}
 - \partial _v A_\alpha + \partial _\alpha A_v  +
\partial _\mu F_{\mu v\alpha }  =0.
\eeqn
\end{subequations}
With regard to the sources we have
\begin{subequations}
\label{sys2}
\beqn
\label{sys2a}
\partial _v \Phi _{\mu v} +A_\mu =0,
\\
\label{sys2b}
\partial _\mu \Phi _{v \alpha } + \partial _\alpha \Phi _{\mu v}
 + \partial _v \Phi _{\alpha \mu }  + F_{\mu v\alpha }=0,
\\
\label{sys2c}
 - \partial _v A_\alpha  + \partial _\alpha A_v  +
\partial _\mu F_{\mu v\alpha } = j_{v \alpha }.
\eeqn
\end{subequations}

System (\ref{sys2}) may be reduced to the correct second-order equations for the
potentials $A_{\mu}  $ and $\Phi _{\mu  v} $. Based on equation (\ref{sys2}a) and
acting on equation (\ref{sys2b}) with the operator $\partial _{\mu}  $ and
substituting $\partial _{\mu}  F_{\mu  v \alpha}  $ into equation (\ref{sys2c}),
we obtain
\begin{equation}
\label{eq26}
\Box\Phi _{v\alpha } = - j_{v\alpha }.
\end{equation}
We get
\begin{equation}
\label{eq27}
\Box A_\mu =  - j_\mu,
\end{equation}
acting on equation (\ref{sys2c}) with $\partial _{v} $, using definition (\ref{eq12}) of open
strings $\left( {j_{\alpha}  =\partial _{\mu}  j_{\mu \alpha}}
\right)$ and based on the antisymmetrical properties
 of tensors $F_{\mu v\alpha}  $
and $\Phi _{\mu \nu}  $.

The field intensity tensor connected with the potential $A_\mu $ is not
appear in manifest form in system (\ref{sys2}). It is in equation (\ref{sys2}) in the
following combination
\begin{equation}
\label{eq28}
\partial _v A_\alpha  - \partial _\alpha A_v \equiv F_{v\alpha } .
\end{equation}
With regard to (\ref{eq28}), one can rewrite equation (\ref{sys2c}) in the form
\begin{equation}
\label{eq29}
 - F_{v \alpha } + \partial _\mu F_{\mu v\alpha } =j_{v\alpha }
\end{equation}
In the absence of the sources it looks as follows
\begin{equation}
\label{eq30}
\partial _\mu F_{\mu v\alpha } =F_{v\alpha }.
\end{equation}
Equation (\ref{eq30}) means that in the case of open string the Maxwell-Kalb-Ramond
field is regarded as a single field realizing the interactions. In the case
when only the interaction of closed strings or point charges are considered,
components of the Maxwell-Kalb-Ramond field can be described separately.
Assuming in (27) that $A_{\mu}  = 0$, we obtain system (\ref{eq11})
describing the Kalb-Ramond field. We get the equation
\begin{equation}
\label{eq31}
\partial _\alpha F_{v\alpha } = j_v ,
\end{equation}
acting on equation (\ref{eq29}) with $\partial \alpha $ and taking into account
equation (\ref{eq12}). We obtain Maxwell system (\ref{eq10})
by considering equation (\ref{eq31})
together with definition (\ref{eq28}) and omitting all other equations in (27)
associated with the string body.

\section{Dirac-K\"{a}hler and Maxwell-Kalb-Ramond fields}

Let us consider now the Dirac-K\"{a}hler field. The name ''Dirac-K\"{a}hler
equation'' (DKE) was introduced in [10]. Its vector form was discovered by
Darwin [11]. Fundamental mathematical properties of the DKE were established
by K\"{a}hler [12]. The DKE has been rediscovered independently in different
mathematical approaches (see, e.g., ref. [13-14] and references therein). A
tensor formulation of DKE is based on the following system
\beqn
\label{eq32}
\nonumber
 \partial _{v} \varphi _{\mu v}  + \partial _\mu \varphi
+ \varphi _\mu  =0,
 \\
 \nonumber
 \partial _\mu \varphi _{v\lambda }  + \partial _\lambda
\varphi _{\mu v}  + \partial _{v} \varphi _{\lambda \mu }
+\partial _\rho \varphi _{\rho \mu v\lambda }  +
\varphi _{\mu v\lambda }  =0,
\\
 - \partial _\mu  \varphi _v  +\partial _v \varphi _\mu +
\partial _\lambda \varphi _{\lambda \mu v}  + \varphi
_{\mu v} = 0,
 \\
\nonumber
 \partial _\mu \varphi _\mu  + \varphi  = 0,
  \\
\nonumber
 \partial _\mu \varphi _{v\lambda \rho }  -\partial _v
\varphi _{\mu \lambda \rho }  + \partial _\lambda \varphi
_{\mu v\rho }  - \partial _\rho \varphi _{\mu v\lambda }
+ \varphi _{\mu v\lambda \rho }  = 0.
\eeqn
The symbol ``$\varphi $'' is used here for all Lorentz covariants: $\varphi
$ is scalar, $\varphi _{\mu}  $ is four-vector, $\varphi _{\mu v}
,\varphi _{\mu v\lambda}  ,\varphi _{\mu v\lambda \rho
} $ are antisymmetric tensors of the 2, 3 and 4 ranks, respectively.
System (\ref{eq32}) has three massless analogs.
One of them is derived from (\ref{eq32}) if
one put in the first two equations the terms $\varphi _{\mu}  $ and
$\varphi _{\mu v\lambda}  $ are equal to zero (see for example
[15]). The second system is obtained if one omit the terms $\varphi _{\mu
v} ,\varphi $ and $\varphi _{\mu v\lambda \rho}  $ in
third, fourth, and fifth equations. The third possibility unifies both
above mentioned massless systems.

One has in the second case
\begin{subequations}
\label{sys3}
\beqn
\label{sys3a}
\partial _v \varphi _{\mu v} + \partial _\mu \varphi +\varphi _\mu =0,
\\
\label{sys3b}
\partial _\mu \varphi _{v\lambda } + \partial _\lambda
\varphi _{\mu v}  + \partial _v \varphi _{\lambda\mu }
 + \partial _\rho \varphi _{\rho \mu v\lambda }  +
 \varphi _{\mu v\lambda }=0,
\\
\label{sys3c}
 - \partial _\mu \varphi _v +\partial _v \varphi _\mu  +
\partial _\lambda \varphi _{\lambda \mu v} = 0,
\\
\label{sys3d}
\partial _\mu \varphi _\mu =0,
\\
\label{sys3e}
\partial _\mu \varphi _{v\lambda \rho } -\partial _v
\varphi _{\mu \lambda \rho }  + \partial _\lambda \varphi
_{\mu v\rho }  - \partial _\rho \varphi _{\mu v\lambda }=0.
\eeqn
\end{subequations}
System (35) is invariant with respect to the gauge transformations
\beq
\begin{split}
 \varphi _\mu  \to \varphi _\mu  + \partial _\mu \Lambda
\left( x \right), \\
 \varphi  \to \varphi - \Lambda \left( x \right),
\end{split}
\label{eq36}
\eeq
(including the case when the sources are present) where the gauge function
$\Lambda \left( x \right)$ satisfies the equation
\begin{equation}
\label{eq35}
\Box\Lambda \left( x \right)=0.
\end{equation}
Taking into account that the function
$\varphi $ in (\ref{sys3}) satisfies the analogical equation
\begin{equation}
\label{eq36_1}
\Box\varphi = 0,
\end{equation}
we can assume $\varphi =
0$ without loss of the generality. In this way we get the following system
\begin{subequations}
\label{sys4}
\beqn
\label{sys4a}
\partial _v \varphi _{\mu v} +\varphi _\mu =0,
\\
\label{sys4b}
\partial _\mu \varphi _{v\lambda } + \partial _\lambda
\varphi _{\mu v}  + \partial _v \varphi _{\lambda \mu }
+ \partial _\rho \varphi _{\rho \mu v\lambda }  +
\varphi _{\mu v\lambda }  = 0,
\\
\label{sys4c}
 - \partial _\mu \varphi _v  + \partial _v \varphi _\mu  +
\partial _\lambda \varphi _{\lambda \mu v}  = 0,
\\
\label{sys4d}
\partial _\mu \varphi _{v\lambda \rho } - \partial _v
\varphi _{\mu \lambda \rho }  + \partial _\lambda \varphi
_{\mu v\rho }  - \partial _\rho \varphi _{\mu v\lambda }=0.
\eeqn
\end{subequations}
We keep also in mind that for $\varphi = 0$ equation
(\ref{sys3d}) follows from (\ref{sys3a}).

According to [3], a massless pseudo-scalar field described by the intensity
tensor $\varphi _{\mu  v \lambda  \rho}  $ represents interaction between
the second-order membranes in the space with the dimensionality $d=
4$. Based on this treatment, we can consider system (39) (or (\ref{sys3})) as a
model for the simultaneous description of the fields realizing interactions
of all the string and membrane types in 4-dimensional space.

If we eliminate from (39) the tensor $\varphi _{\mu v\lambda  \rho}  $
(equation (\ref{sys4d}) in this case follows from (\ref{sys4b})),
we get system
(26), i.e. the description of the Maxwell-Kalb-Ramond massless field. With
regard to the sources, we have system (27).

System (35) is formed on the basis of the following linking scheme
\begin{equation}
\label{eq38}
\unitlength= 0.20mm
\begin{picture}(0,+90)(0,0)
\special{em:linewidth 0.4pt} \linethickness{0.4pt}
\put( 0,90){$(0,0) $}
\put(-70,0){$(0,1) $}
\put(+80,0){$(1,0) $,}
\put(-2,+40){$({1\over 2},{1\over 2} ) $}
\put(-2,-40){$({1 \over 2},{1\over 2} )'$}
\put( 0,-90){$(0,0)' $}

\put(-50,+15){\line(2,+1){45}}
\put(-50,-10){\line(2,-1){45}}
\put(+90,+15){\line(-2,+1){45}}
\put(+90,-10){\line(-2,-1){45}}
\put(+20,+60){\line(0,+1){20}} \put(+20,-50){\line(0,-1){20}}

\end{picture}
\vspace{2cm}
\end{equation}
where $\left( {0,0} \right)^\prime $ is pseudoscalar representation
identical to  the tensor $\varphi _{\mu v\lambda \rho } $.
Let us number the representations entering into (\ref{eq38}) as follows:
$(0,0)\sim 1,~(0,0)^\prime \sim 2,~(\frac{1}{2},\frac{1}{2})\sim 3,~
 (\frac{1}{2},\frac{1}{2})^\prime \sim 4,~(0,1)\sim 5,~
 (1,0)\sim 6,$. Then in the Gel'fand-Yaglom basis,
 we obtain using all the standard
requirements of a RWE theory the following expressions for matrices
$\gamma_4$ and $\gamma_0$
\begin{equation}
\label{eq39}
\gamma _4 =\left( {{\begin{array}{*{20}c}
 {C^0} \hfill & \hfill \\
 \hfill & {C^1 \times I_3 } \hfill \\
\end{array} }} \right),~C^0 =\left(
{{\begin{array}{*{20}c}
 0 \hfill & 0 \hfill & 1 \hfill & 0 \hfill \\
 0 \hfill & 0 \hfill & 0 \hfill & 1 \hfill \\
 1 \hfill & 0 \hfill & 0 \hfill & 0 \hfill \\
 0 \hfill & 1 \hfill & 0 \hfill & 0 \hfill \\
\end{array} }} \right),~C^1=\frac{1}{\sqrt 2
} \left( {{\begin{array}{*{20}c}
 0 \hfill & ~~0 \hfill & 1 \hfill & ~~1 \hfill \\
 0 \hfill & ~~0 \hfill & 1 \hfill & { - 1} \hfill \\
 1 \hfill & ~~1 \hfill & 0 \hfill & ~~0 \hfill \\
 1 \hfill & { - 1} \hfill & 0 \hfill & ~~0 \hfill \\
\end{array} }} \right),
\end{equation}
\begin{equation}
\label{eq40}
\gamma _0 =\left( {{\begin{array}{*{20}c}
 {O_2 } \hfill & \hfill & \hfill \\
 \hfill & {I_8 } \hfill & \hfill \\
 \hfill & \hfill & {O_6 } \hfill \\
\end{array} }} \right) =\left( {{\begin{array}{*{20}c}
 {O_2 } \hfill & \hfill & \hfill & \hfill \\
 \hfill & {I_2 } \hfill & \hfill & \hfill \\
 \hfill & \hfill & {I_6 } \hfill & \hfill \\
 \hfill & \hfill & \hfill & {O_6 } \hfill \\
\end{array} }} \right).
\end{equation}
Omitting the representations
$\left( {0,0} \right)$ and $\left( {0,0} \right)^{\prime} $ in
the linking scheme of (\ref{eq38}), we get at RWE (13)
with linking scheme (\ref{eq21}) and
matrices $\gamma _{4} $ (\ref{eq22}) and $\gamma _{0} $ (\ref{eq23}),
i.e. the theory
describing, as has been noted previously, the Maxwell-Kalb-Ramond (photon +
notoph) field realizing interactions of the open strings (first-order
membranes) in the space with the dimensionality $d   =   4$.

Therefore, this theory is a particular case  of the Dirac-K\"{a}hler system.

\section{Conclusion}

So, we see that both cases of massless vector field (with helicity values
$\pm $ 1 and with zero value) must be considered on equal grounds from the
point of view of RWE in the Gel'fand-Yaglom approach. In fact, the
description of the photon-notoph system  is the direct consequence of its
consideration in the above mentioned theory. It is also evident that the
Kalb-Ramond field equations and the photon-notoph equations are particular
massless cases of the Dirac-Kahler equation. It opens new possibilities
for applications of the Dirac-Kahler field in the string theory, because the
Kalb-Ramond field appears in the effective low-energy field theory
derived from relativistic strings(see for example [16]).

\end{document}